# Beyond empirical models: Discovering new constitutive laws in solids with graph-based equation discovery

Hao Xu[1,2], Yuntian Chen[1,3,*], and Dongxiao Zhang[1,4,*]

[1] Zhejiang Key Laboratory of Industrial Intelligence and Digital Twin, Eastern Institute of Technology, Ningbo, Zhejiang 315200, China

[2] Department of Electrical Engineering, Tsinghua University, Beijing 100084, P. R. China.

[3] Ningbo Institute of Digital Twin, Eastern Institute of Technology, Ningbo, Zhejiang 315200, P. R. China

[4] Institute for Advanced Study, Lingnan University, Tuen Mun, Hong Kong

[*] Corresponding authors.

Email address: ychen@eitech.edu.cn (Y. Chen); dzhang@eitech.edu.cn (D. Zhang).

**Abstract**

Constitutive models are fundamental to solid mechanics and materials science, underpinning the quantitative description and prediction of material responses under diverse loading conditions. Traditional phenomenological models, which are derived through empirical fitting, often lack generalizability and rely heavily on expert intuition and predefined functional forms. In this work, we propose a graph-based equation discovery framework for the automated discovery of constitutive laws directly from multisource experimental data. This framework expresses equations as directed graphs, where nodes represent operators and variables, edges denote computational relations, and edge features encode parametric dependencies. This enables the generation and optimization of free-form symbolic expressions with undetermined material-specific parameters. Through the proposed framework, we have discovered new constitutive models for strain-rate effects in alloy steel materials and the deformation behavior of lithium metal. Compared with conventional empirical models, these new models exhibit compact analytical structures and achieve higher accuracy. The proposed graph-based equation discovery framework provides a generalizable and interpretable approach for data-driven scientific modelling, particularly in contexts where traditional empirical formulations are inadequate for representing complex physical phenomena.

**Introduction**

Constitutive laws serve as fundamental elements in solid mechanics, establishing the relationship between kinematic measures and static quantities to characterize material-specific behavior. Unlike conservation principles and kinematic relations, which are derived from first principles and regarded as axiomatic foundations, constitutive models encapsulate empirical descriptions of material responses to external stimuli. Accordingly, they are typically established through phenomenological approaches, guided by systematic experimentation and theoretical generalization, to characterize nonlinear behaviors across varying conditions(*1*). The accuracy and generality of constitutive models are critical for the reliability of mechanical analysis, directly influencing both theoretical developments and practical applications in computational mechanics and materials engineering.

Traditionally, the development of constitutive laws has relied on empirical or semiempirical approaches (*2*). Empirical models are primarily based on fitting experimental data, often without explicit consideration of underlying physical mechanisms, resulting in good performance within calibrated ranges but limited generalizability beyond those conditions (*3*, *4*). Semiempirical models integrate theoretical principles, such as invariance requirements and thermodynamic consistency, to enhance physical interpretability, but remain reliant on experimental calibration(*5*, *6*). In general, despite their widespread application, phenomenological models exhibit inherent limitations, including dependency on domain expertise for formulation and constrained generalizability, and often fail to accurately represent complex, history-dependent behaviors characteristic of advanced materials.

The emergence of artificial intelligence (AI) and data-driven approaches has led to the introduction of alternative frameworks for constitutive modelling(*7*, *8*). Recent research in this field has advanced primarily in two directions. The first direction focuses on model-free strategies, wherein black-box machine learning models directly map experimental data such as strain–stress relationships without explicitly formulating constitutive laws (*9–11*). The second direction focuses on parameter-learning methods, where neural networks are employed to learn parameters within predefined constitutive models, which are subsequently integrated into numerical solvers for computational analysis (*12–14*). While these approaches are capable of approximating complex and nonlinear material behaviours, their black-box nature hinders interpretability and constrains their applicability across diverse material systems. Moreover, their strong dependence on data limits their effectiveness in scenarios where high-quality data are scarce or costly to acquire.

Accordingly, symbolic and sparse regression have gained attention as promising tools for the data-driven discovery of interpretable constitutive relationships. However, their conceptual appeal is often offset by practical limitations that restrict their effectiveness in real-world applications (*15–19*). Symbolic regression explores a vast combinatorial space of mathematical expressions but often produces overly complex or physically inconsistent formulations, particularly for materials with strong nonlinear or history-dependent responses (*20–22*). This stems in part from its use of binary tree representations, which are low in information density and prone to generating fragmented, hard-to-interpret structures. Sparse regression, although computationally efficient, depends on predefined libraries of candidate terms, which limits its discovery potential when prior knowledge of the governing functional forms is incomplete or absent (*23–26*). Although recent efforts have succeeded in recovering governing equations from observational data, these approaches have focused primarily on differential equations, where the complexity arises from differential terms rather than the underlying functional structure (*27–30*).

Another common but critical challenge faced by these approaches is their inability to accommodate the parametric nature inherent in constitutive laws. In constitutive modelling, it is common practice to incorporate material-specific parameters that can be calibrated from experimental data, enabling the model to adapt across different materials. However, existing methods are generally incapable of discovering equations with undetermined parameters. Currently employed symbolic techniques are limited to generating fixed-coefficient equations by their tree-based representational forms and cannot effectively encode expressions with free parameters (*31–34*). In addition, existing methods often rely on homogeneous, single-source datasets, which increase the risk of overfitting under limited data availability, and hinder the discovery of general constitutive relationships applicable across diverse materials and loading conditions.

In this work, we propose a data-driven framework, called graph-based equation discovery (GraphED), for discovering constitutive laws in solids. Based on the concept of "equations as graphs", this framework establishes a graph-based representation of mathematical expressions, where operators and variables are encoded as nodes within a directed graph, and their relationships are defined by edges (Fig. 1A). Each edge carries a parameter (i.e., edge feature) that represents either a constant value or a learnable coefficient, capturing the parametric dependency. Compared with conventional tree-based representations commonly used in symbolic regression, the graph-based representation offers several advantages: (1) By allowing node sharing and flexible edge connectivity, the graph representation enables a more compact encoding of mathematical expressions, particularly in scenarios involving repeated additive or multiplicative operations and complex symbolic dependencies. (2) Computational order is explicitly defined through directed edges. (3) Parameters are embedded directly as edge features, which avoids the need for additional parameter nodes and encodes material-specific parameters.

To prevent the generation of overly complex or implausible expressions, graph templates are introduced, which define structural patterns for different operators by imposing structural constraints on the subgraphs associated with specific operators. Genetic programming, along with specific graph crossover and mutation is employed for optimization. In case studies, new constitutive models for rate-dependent alloy steel materials have been discovered from published real-world experimental data, which outperforms existing empirical models in terms of both accuracy and structural compactness. Furthermore, for the deformation of lithium, which cannot be described by conventional models, the proposed method has identified new constitutive laws directly from experimental data with high accuracy. These results underscore the potential of the proposed framework for the discovery of constitutive models in scientific applications.

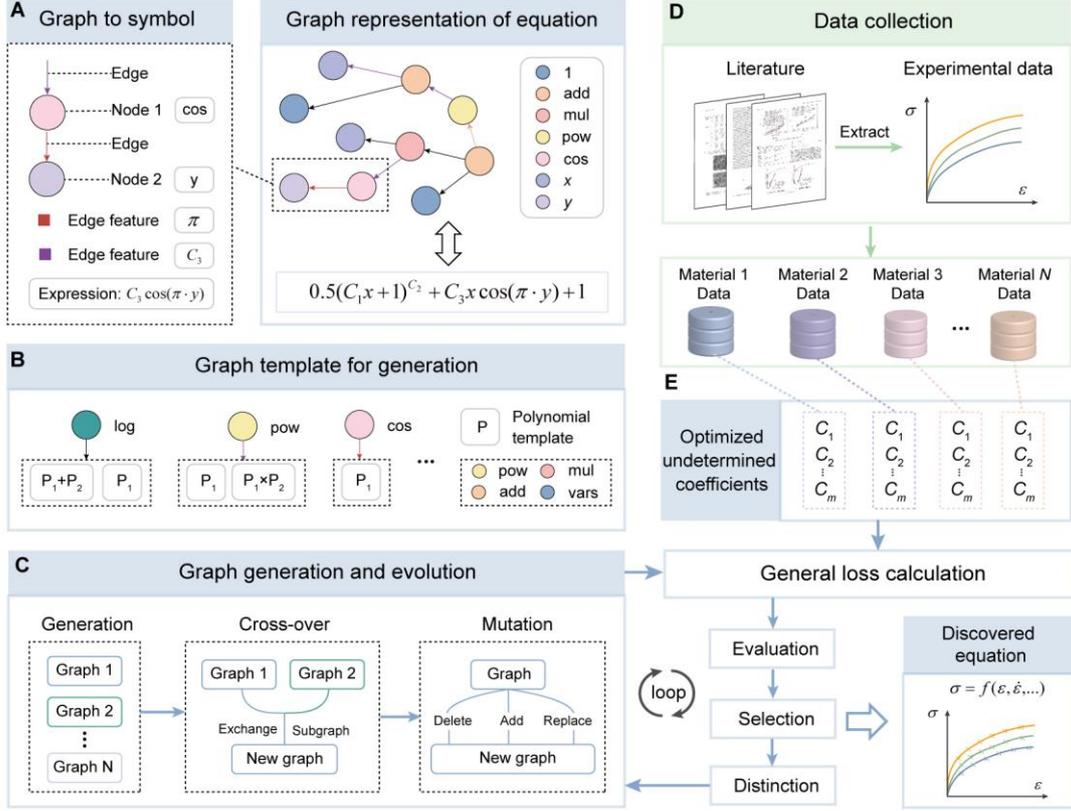

**Fig. 1. The architecture of the proposed graph-based equation discovery (GraphED) framework for the discovery of constitutive laws.** **(A)** Graph-based representation of equations, where nodes correspond to operators and variables, and edges represent their connections. Edge features encode parametric information, including constants or undetermined material-specific coefficients. **(B)** Diagram of the graph template, which imposes structural constraints on the subgraphs associated with specific operators. **(C)** Graph evolution process, including generation, crossover, and mutation operations. **(D)** Experimental data are collected from the literature to construct a diverse, multisource dataset. **(E)** Optimization of the most suitable equation from the datasets.

## Results
### The proposed graph-based equation discovery framework
The core of the proposed graph-based equation discovery framework is "equation as graph". A graph is a versatile data structure used to model entities and their relationships, where nodes represent objects and edges define their interactions. Owing to their expressive capacity, graphs are widely applied in the analysis of large-scale, interdisciplinary problems, including social networks(*35*), physical systems(*36*), and chemical molecular characterization(*37*). As depicted in Fig. 1A, mathematical equations are encoded as directed acyclic graphs, where nodes represent operators and variables, and edges define their connections. All nodes in the graph are constrained to have an in-degree of one, whereas the out-degree varies according to the type of node. Owing to their associative properties, operators such as *add* and *mul* can have multiple outgoing edges, which guarantees representation compactness. To handle material-specific parameters in constitutive models, each graph edge is assigned a feature corresponding to either a constant or an undetermined parameter, which encodes the internal coefficient of the corresponding subexpression. During graph traversal, the edge feature is typically interpreted as a multiplicative factor applied to the sub-expression represented by the downstream node (Fig. 1A). Specifically, for operators such as *pow* and *log*, where the exponent or base must be explicitly specified, the required values are encoded within the edge features. As a result, only a single edge is needed to connect the operator node to its operand subexpression.

The fundamental distinction between conventional tree-based and graph-based representations lies in their structural flexibility and expressiveness. While tree structures offer simplicity and implicit control over computational order, they are prone to redundancy and limited adaptability when they represent complex expressions such as polynomials or deeply nested operations, which can lead to rapid increases in tree depth and node count, thereby reducing optimization efficiency. Moreover, tree-based expressions struggle to accommodate dynamic parameters and impose complex constraints, which are essential for the discovery of constitutive laws. In contrast, the proposed graph-based representation achieves more compact and information-dense expressions through node sharing and flexible edge connections, making it particularly effective in scenarios involving intricate symbolic dependencies. By introducing edge features, the graph-based representation naturally incorporates parametric dependencies and symbolic constraints, thereby enhancing semantic expressiveness and rendering it more suitable for scalable symbolic modelling in data-driven scientific discovery.

In this framework, genetic programming is utilized for optimization through a loop of "generation-evaluation-evolution". For the generation process, graph templates are introduced to define a set of structured rules and configurable subgraph patterns that govern the arrangement of operators and operands within the constructed graph (Fig. 1B). A basic subgraph template *P* is defined, comprising only *add*, *mul*, *pow*, and nodes of variables, with constraints placed on the depth of the subgraph structure. For other operator types (e.g., *log*, *exp*, *cos*), permissible subgraphs are predefined based on the template *P*. Notably, graph templates are not static, they can be adapted and customized to accommodate domain-specific requirements and problem contexts. During graph generation, candidate graphs are assembled from randomly generated subgraphs that adhere to these predefined templates. This modular approach mitigates the risk of generating overly complex structures, thereby constraining the search space to structurally meaningful regions. After graph generation, graph crossover and mutation are defined (Fig.1C), which is detailed in the Materials and Methods.

The generated graphs are evaluated using experimental datasets collected from the literature across various materials (Fig. 1D). Since constitutive laws typically involve material-specific undetermined parameters, the L-BFGS optimization algorithm is employed to fit these parameters individually for each material dataset. The mean error computed across all datasets is used as the evaluation metric for each graph, thereby promoting the discovery of equations with generalizability. On this basis, a selection process is implemented whereby the top 50% of graphs with the lowest finesses are retained, while the remaining candidates are replaced with newly generated random graphs to maintain population diversity. To further reduce the risk of premature convergence to local optima, an extinction strategy is introduced where all but a small subset of top-performing graphs is eliminated and replaced with randomly generated graphs. Throughout the evolutionary process, the framework progressively converges toward optimal constitutive equations (Fig. 1E).

To validate the effectiveness of the proposed framework, we apply it to the discovery of new constitutive models for strain-rate effects in alloy steels and for the deformation behavior of lithium metal. These problems involve complex material behavior, including dynamic strengthening, strain hardening, rate dependence, and thermal sensitivity, which poses a challenge to conventional empirical modelling approaches.

**Discovery of the dynamic increase factor constitutive model**

The dynamic increase factor (DIF) is a critical concept in material modelling that quantifies the rate-dependent enhancement of material strength under dynamic loading conditions. Specifically, DIF is defined as the ratio between a material property measured under dynamic loading and its corresponding quasi-static value, which is defined as:

$$\text{DIF} = \frac{\sigma_{\text{dynamic}}}{\sigma_{\text{static}}}. \tag{1}$$

In this equation, $\sigma_{\text{dynamic}}$ and $\sigma_{\text{static}}$ represent the dynamic and static strengths, respectively. DIF constitutive models are essential for extending conventional static material descriptions to scenarios involving strain-rate sensitivity. Most existing models adopt an empirical form in which DIF is expressed as a function of the strain rate. (Supplementary Information S1.1).

In this study, we aim to discover a generalized constitutive model for DIF directly from experimental data extracted from the literature through the proposed GraphED framework. A total of 408 groups of strain rate-DIF data points for 40 different materials under varying strain rates were compiled from published studies with a primary focus on alloy steels (Table S1). As shown in Fig. 2A, the collected data cover a wide range of strain rates, from $10^{-4}$ to $10^4$, ensuring comprehensive representation across different loading conditions. This broad coverage enhances the applicability of the discovered model across diverse strain-rate regimes. For the equation discovery process, the operator set comprises *add*, *mul*, *exp*, *pow*, and *log*, with the strain rate as the independent variable and DIF as the target variable. The initial population size was set to 300, and optimization was performed over 150 epochs. The number of undetermined parameters in each candidate equation was restricted to a maximum of two. The detailed experimental settings are provided in Supplementary Information S1.2.

The optimization trajectory and corresponding results are shown in Fig. 2B. During the optimization process, a notable decrease in the loss values is observed, indicating effective convergence. Among the final top five equations, the leading three exhibit identical loss values and are mathematically equivalent, despite being expressed in different forms. This consistency highlights the stability of the algorithm and reinforces the reliability of the discovered models. The optimal equation is written as:

$$DIF = [(C_1 \cdot \dot{\varepsilon} + 1)^{-2}]^{C_2} \qquad (2)$$

where $\dot{\varepsilon}$ is the strain rate and $C_1$ and $C_2$ are the undermined material parameters. This concise form exhibits a generalized power-exponential form with unknown parameters embedded in the exponent, reflecting both simplicity and interpretability. In contrast, the suboptimal equations display both higher loss values and greater structural complexity. This demonstrates the ability of our method to generate and identify intricate equation forms with free parameters, which remains challenging for traditional tree-based equation discovery approaches.

To evaluate the accuracy of the discovered constitutive model, a comparative analysis is conducted against several classical DIF constitutive models, including the Cowper-Symonds (C-S) model(*38*), Johnson-Cook (J-C) model(*39*), Huh & Kang (H-K) model(*40*), and Rule & Jones (R-J) model(*41*). Descriptions of these models are provided in Supplementary Information S1.2. As shown in Fig. 2C, the proposed model achieves the lowest mean squared error (MSE) of $6.9 \times 10^{-4}$ and the highest coefficient of determination ($R^2$) of 0.980. Notably, even the best-performing existing model, the H-K model, which also contains two material-specific parameters, yields an MSE of $1.2 \times 10^{-3}$, nearly twice that of the discovered model. For the comparison of model generalizability, Fig. 2D presents the performance of each model across four representative materials that exhibit distinct rate-dependent behaviors within different strain-rate regimes. The discovered model demonstrates robust performance across the entire strain-rate spectrum, including extreme conditions such as very low ($10^{-4}$) and very high ($10^3$) strain rates. In contrast, the C-S and J-C models exhibit substantial errors throughout all strain-rate ranges. Although the R-J and H-K models perform well at high strain rates, their fitting accuracy deteriorates in the low strain rate region (<1). A comparison between these models on more materials is provided in Fig. S2.

These observations highlight the limitations of traditional empirical models: they are typically derived under specific experimental assumptions. While such models may perform adequately within their calibrated ranges and specific materials, they often lack generalizability. Moreover, constrained by human intuition, existing models predominantly rely on linear relationships or simple nonlinear functions, thereby limiting their expressive capacity. In contrast, the proposed GraphED framework can discover free-form constitutive laws directly from multisource experimental data, eliminating the need for manual derivation. The resulting model not only demonstrates excellent fitting performance across diverse materials and strain rate ranges but also achieves a balance between parsimony and accuracy. Even with the same number of material-specific parameters, the data-driven constitutive law exhibits superior predictive capability.

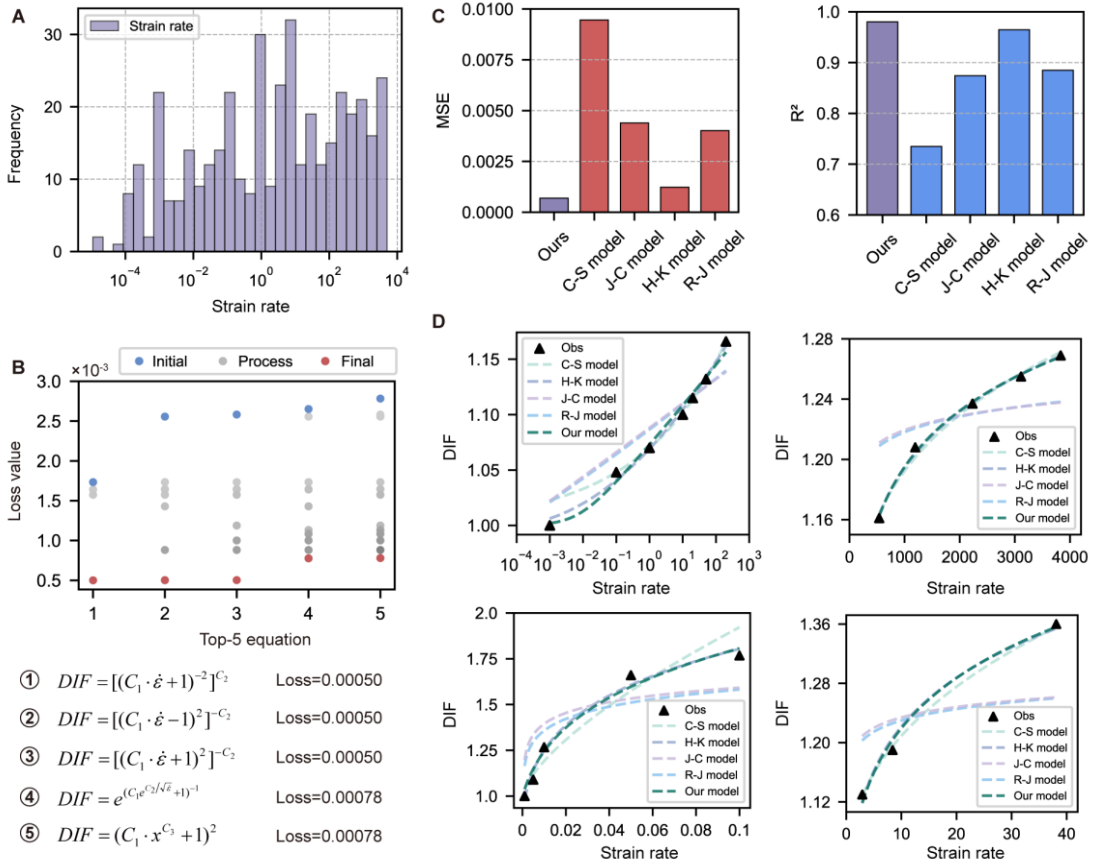

**Fig. 2. Data-driven discovery of dynamic increase factor (DIF) constitutive model.**
**(A)** Distribution of the strain rate in the datasets collected from the literature. **(B)** Optimization trajectory and the top five discovered equations with their corresponding loss values. The darker grey color indicates larger optimization epochs. **(C)** Comparison of MSEs (left) and $R^2$ values (right) between the discovered constitutive model and four conventional empirical models. **(D)** Observed DIF curves and fitted DIF curves generated by different constitutive models for four materials across varying strain rate ranges.

**Discovery of the strain hardening constitutive model**

Strain hardening is a fundamental material phenomenon wherein a material exhibits an increase in flow stress with progressive plastic deformation. This behavior arises from the accumulation and interaction of dislocations within the microstructure, which impedes further plastic flow and thus enhances the resistance of the material to deformation. The strain hardening constitutive model explicitly incorporates this dependency, describing the evolution of stress as a function of plastic strain. Here, we focus on the constitutive law for isotropic hardening, which is important for simulating metal formation and structural crashworthiness.

For the data-driven discovery of constitutive equations, a total of 64 plastic strain–stress curves from 12 different materials under varying strain rates were collected from the literature (Table S1). As shown in Fig. 3A, the dataset covers a wide range of strains, from $10^{-4}$ to $10^0$. In total, the dataset consists of 1,314 data points. For the equation discovery process, the operator set included *add*, *mul*, *exp*, *pow*, and *log* functions. The true plastic strain was used as the independent variable, and the true plastic stress was set as the target variable. The initial population size was fixed at 300, and optimization was performed over 150 epochs. To maintain model simplicity and interpretability, the number of unknown parameters in each candidate equation was limited to a maximum of three. The detailed experimental settings are provided in Supplementary Information S1.2.

The optimization trajectory and corresponding results are shown in Fig. 3B, which demonstrate the effectiveness of the optimization. The final top five equations exhibit diverse structural forms, which underscores the broad yet effective exploration capability of the proposed method. Notably, all the discovered equations demonstrate the ability to capture asymptotic behavior, a critical feature for accurately describing the strain hardening saturation phenomenon. Although the loss values of the top two equations are very close, their mathematical structures differ substantially. Based on an assessment of mathematical behavior and physical consistency, the second-best equation was selected as the final model, which is written as:

$$\sigma = C_1 e^{\left(\frac{1}{C_2 \cdot \varepsilon + C_3}\right)} \tag{3}$$

Compared with the rational-form hardening expression, the exponential-form model provides superior mathematical smoothness, avoids singularities within the plastic strain domain, and more accurately captures the trend of strain hardening characterized by a rapid initial increase followed by gradual saturation.

To evaluate the accuracy of the discovered constitutive equation, a comparative analysis was conducted against several classical strain hardening models, including the Hollomon(*42*), Voce(*43*), Ludwik(*44*), Hartley and Srinivasan (H-S)(*45*), Ludwigson(*46*), and Bargar models(*47*) (Fig. 3C). Detailed descriptions of these models are provided in Supplementary Information S1.1. Since strain hardening behavior has been extensively studied and generally follows a relatively simple trend, most classical empirical models achieve good fitting performance, with $R^2$ values typically exceeding 0.97. However, these models usually require four unknown material parameters to maintain high accuracy, thereby compromising model simplicity. In contrast, the discovered equation achieves comparable, or even superior, performance when only three material-specific parameters are used, offering a better balance between accuracy and parsimony. Notably, although exponential forms appear in some classical models, such as the Ludwigson and Voce models, their exponents are typically restricted to simple linear combinations of the true plastic strain, reflecting the limitations imposed by human intuition in empirical modelling. In contrast, our discovered equation implies that simply applying an inverse form of the plastic strain leads to improved modelling performance, highlighting the advantage of data-driven discovery over traditional empirical approaches. Fig. 3D presents the plastic strain–stress curves of four representative materials, each of which exhibited distinct strain hardening characteristics. While existing empirical models generally provide good fitting performance, the discovered model demonstrates superior accuracy in cases with pronounced hardening saturation. This result indicates that the discovered equation possesses strong generalizability and can effectively capture a wider range of hardening behaviors across different materials. A comparison between these models on more materials is provided in Fig. S2.

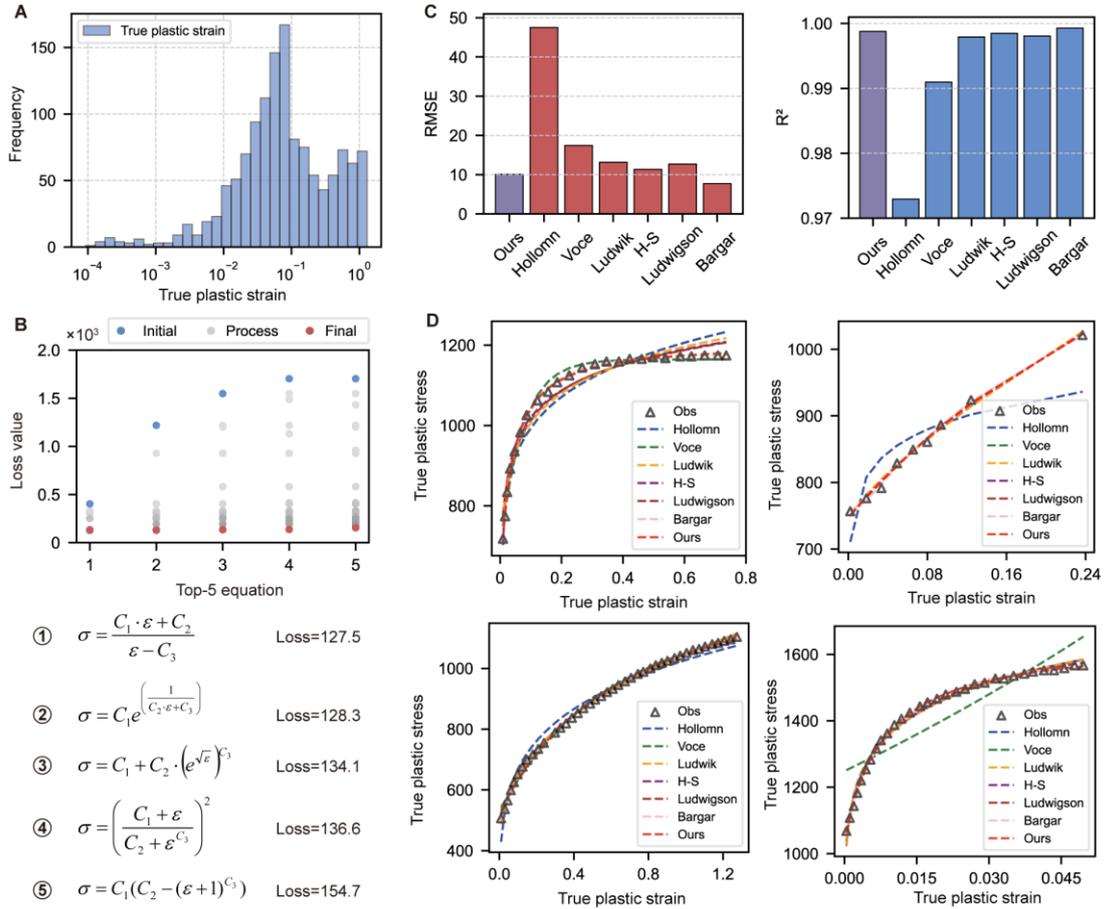

**Fig. 3. Data-driven discovery of strain hardening constitutive model. (A)** The distribution of true plastic strain in the datasets collected from the literature. **(B)** Optimization trajectory and the top five discovered equations with their corresponding

loss values. The darker grey color indicates larger optimization epochs. **(C)** Comparison of MSEs (left) and $R^2$ values (right) between the discovered constitutive model and six conventional empirical models. The white notation is the number of undetermined material-specific parameters in the corresponding model. **(D)** Observed and fitted plastic strain–stress curves generated by different constitutive models for four materials.

**Construction of an integrated rate-dependent constitutive model**

Building on the independently discovered constitutive models for dynamic increase factor and strain hardening, we aim to construct an integrated rate-dependent constitutive relation that describes the true plastic stress as a joint function of the true plastic strain and strain rate. This integrated model is designed to provide a comprehensive and physically consistent description of rate-dependent plastic deformation across a wide range of loading conditions, which is derived as:

$$\sigma_{dynamic} = \sigma_{st} \cdot DIF = C_1^{st} e^{1/(C_2^{st} \cdot \varepsilon + C_3^{st})} \cdot [(A \cdot \dot{\varepsilon} + 1)^{-2}]^{0.5B} . \tag{4}$$

Here, $C_1^{st}$, $C_2^{st}$, and $C_3^{st}$ represent the material-specific parameters of the discovered strain hardening model under the quasi-static state. $A$ and $B$ represent the material-specific parameters in the discovered DIF constitutive model. The constructed integrated model enables the prediction of dynamic plastic stress–strain curves under arbitrary strain rates based on the quasi-static curves obtained from tensile tests. Fig. 4A shows the model performance for two strain rate materials. Across a wide range of strain rates spanning several orders of magnitude, the model consistently achieves high-accuracy predictions of the corresponding plastic stress–strain responses. These results validate the effectiveness of the discovered model in capturing rate-dependent plastic behavior.

Furthermore, a comparative analysis is conducted against the widely used empirical integrated rate-dependent constitutive model, the Johnson–Cook (J-C) model(*39*), which accounts for the effects of strain hardening and strain rate sensitivity. The general form of the J–C model is as follows:

$$\sigma = [A + B \cdot \varepsilon^n] \cdot \left[1 + C \cdot \ln\left(\frac{\dot{\varepsilon}}{\dot{\varepsilon}_0}\right)\right] \tag{5}$$

where $A$, $B$, $n$, and $C$ represent the material parameters and $\dot{\varepsilon}_0$ is the strain rate of the quasi-static state. Since all the experimental data were collected at room temperature, the influence of temperature is ignored in this analysis. The performances of the discovered model and J-C model are evaluated across experimental datasets from ten different materials, with the overall mean squared error (MSE) and coefficient of determination ($R^2$) computed for assessment. Although the J-C model exhibits strong overall performance (MSE = 1108.9, $R^2$ = 0.955), the discovered model achieves higher accuracy, with an MSE of 646.4 and an $R^2$ of 0.974, reducing the prediction error to nearly half that of the J–C model.

For a more detailed comparison, a challenging case, high-strength steel Q460JSC, is selected from the dataset(*48*). This material exhibits dynamic response characteristics at high strain rates that differ significantly from those of conventional materials. As shown in Fig. 4B and 4C, although the J–C model fits the quasi-static data reasonably well, it fails to accurately extrapolate to high strain rates, resulting in substantial prediction errors. In contrast, the discovered model demonstrates superior adaptability, achieving close agreement with the observed data across the entire strain-rate range, despite minor deviations at extremely high strain rates. These results highlight the generalizability of the proposed integrated model across different conditions.

Further analysis revealed that, in certain complex cases, a degree of coupling may exist between strain hardening and strain rate sensitivity. To address this, we propose a preliminary modification: replacing the coefficients $C_{st}^1$ and $C_{st}^2$ in Equation (4) with $C_{st}^1 - \lambda_1 \times \text{DIF}$ and $C_{st}^2 - \lambda_2 \times \text{DIF}$, respectively. This modified model exhibits improved performance, particularly under high strain rate conditions (Fig. 4C). Notably, the potential coupling effects between hardening and rate sensitivity are complex, and the proposed modification remains a preliminary attempt. Nevertheless, these findings demonstrate that the data-driven discovered constitutive model not only aligns with experimental data but also exhibits strong expandability.

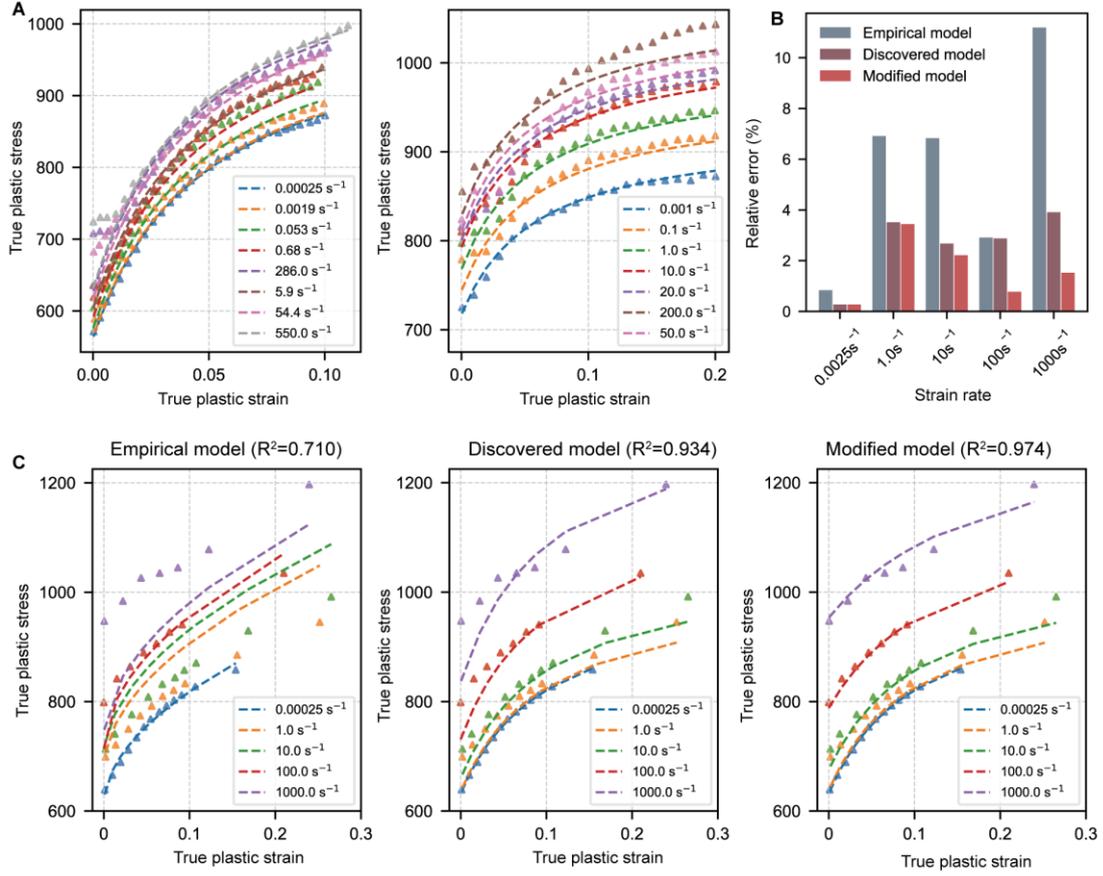

**Fig. 4. The performance of the constructed integrated constitutive model. (A)** Observation (triangle scatter) and prediction (dotted line) of plastic strain–stress curves by the constructed integrated constitutive model. **(B)** Comparison of the relative error under different strain rates among the empirical model, the model constructed from the discovered equations, and the modified model. **(C)** Comparison of the predicted plastic strain–stress curves under different strain rates for high-strength steel Q460JSC between the empirical model, the model constructed from the discovered equations, and the modified model.

**Discovery of the rate-dependent constitutive model for lithium**

In this section, we extend our framework to address a more challenging problem: the discovery of a rate-dependent constitutive model for the deformation and creep behavior of lithium metal. Although lithium plays a critical role in advanced technologies, such as battery systems and energy storage devices, its mechanical behavior remains relatively underexplored. Given the limitations of conventional power–law flow models in capturing the complex deformation and creep behavior of lithium, there is a critical need for a unified, rate-sensitive constitutive model that can accurately describe its mechanical response over broad temperature and strain-rate ranges. The objective of this study is to discover a rate-dependent constitutive model that accurately characterizes the plastic flow behavior of lithium across different stress conditions. The mechanical dataset is compiled from diverse experimental studies reported in the literature (*49*, *50*), which includes the normalized plastic strain rate ($\dot{\gamma}/\dot{\bar{\varepsilon}}$) and applied stress ($\sigma$) values under various temperatures and strain rates. In this study, $\dot{\gamma}$ represents the plastic strain rate and $\dot{\bar{\varepsilon}}$ represents the equivalent strain rate.

The influence of temperature is examined first. In the dataset, the flow curves of lithium metal were recorded at a fixed strain rate of $3\times10^{-5}$ s$^{-1}$ at five different temperatures: 198 K, 248 K, 273 K, 298 K, and 348 K. The dataset comprises a total of 119 data points. For the equation discovery process, the operator set included *add*, *mul*, *exp*, *pow*, and *log* functions. The initial population size was set to 300, and optimization was performed over 150 epochs. The number of unknown parameters allowed in each candidate equation was restricted to a maximum of three. Preliminary equation discovery of the experimental data revealed that the normalized plastic strain rate exhibits an approximately linear relationship with the applied stress across different temperatures (Table S2). Therefore, we performed the proposed GraphED framework separately on the slope and intercept of the linear relationships for the plastic flow curves to discover their underlying temperature dependences, as detailed in Supplementary Information S1.3. The final discovered equation is as follows:

$$\dot{\gamma}/\dot{\bar{\varepsilon}} = 0.543\sigma + 0.83 \big/ (1+e^{(-0.006T)^{2.51}})^{6.66} \qquad (6)$$

Interestingly, in contrast to conventional power-law models with an Arrhenius-type thermal component, the discovered constitutive law has a distinct structure. This equation combines a direct linear dependence on stress with a nonstandard evolution term involving an exponential function of temperature *T*. This equation suggests that shear–tensile coupling is driven primarily by stress while being concurrently modulated by an independent temperature-dependent mechanism, potentially associated with microstructural rearrangement and thermal softening. The decoupling of the stress response from the evolutionary term provides greater flexibility in capturing the progressive nature of shear-dominated deformation under dynamic conditions, representing a notable departure from traditional multiplicative constitutive models.

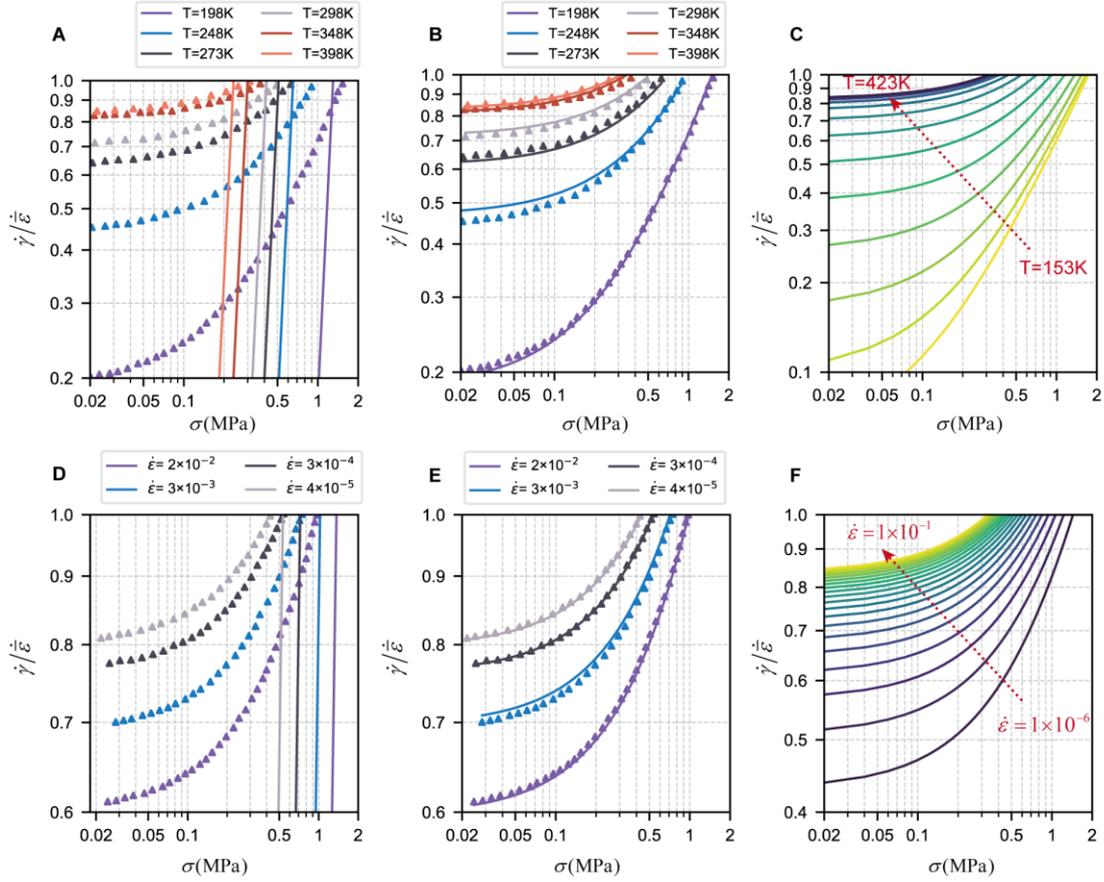

**Fig. 5. The performance of the discovered rate-dependent constitutive model for lithium. (A) & (B)** Observation (triangle scatter) and prediction (dotted line) of plastic flow curves under different temperatures by the (A) empirical model and (B) discovered model. **(C)** Generalization of the discovered model to broader temperature regions. **(D) & (E)** Observation and prediction of plastic flow curves under different strain rates by the (D) empirical model and (E) discovered model. **(F)** Generalization of the discovered model to broader strain rate regions.

As shown in Fig. 5A and Fig. 5D, the conventional empirical models exhibit steep, nearly vertical transitions at intermediate stress levels, resulting in substantial deviations from the experimentally observed gradual increase in the normalized strain rate. This behavior underscores a key limitation of classical creep formulations, which often rely on power-law mechanisms and therefore fail to capture the smooth, continuous evolution of deformation across varying thermal conditions. In contrast, the model discovered through the proposed GraphED framework demonstrates excellent agreement with the experimental data across all temperature regimes, achieving a relative error of only 1.86% (Fig. 5B). Furthermore, Fig. 5C illustrates the extrapolation capability of the discovered model beyond the experimental range. The results show that it successfully captures the underlying deformation trends, indicating strong generalizability and robustness.

Next, the influence of the strain rate was examined. In the dataset, flow curves of lithium metal were recorded at a fixed temperature of 273 K for four different strain rates: $2\times10^{-2}$ s$^{-1}$, $3\times10^{-3}$ s$^{-1}$, $3\times10^{-4}$ s$^{-1}$, and $4\times10^{-5}$ s$^{-1}$, comprising a total of 199 data points. The settings for equation discovery were identical to those used in the temperature-dependent analysis. To better capture the underlying relationships, a logarithmic transformation was applied to the strain rate data prior to model discovery. Preliminary equation discovery of the experimental data revealed that the normalized plastic strain rate also exhibits an approximately linear relationship with the applied stress across different strain rates (Table S4). The detailed discovery process is provided in Supplementary Information S1.3. The final discovered equation is as follows:

$$\dot{\gamma}/\dot{\bar{\varepsilon}} = (0.77 - 0.008 \log \dot{\varepsilon})\sigma + 0.956 e^{0.8/\log \dot{\varepsilon}} \tag{7}$$

In conventional power-law constitutive models for lithium metal, the influence of the strain rate is not explicitly incorporated (Fig. 5D). In contrast, the equation discovered in this study reveals a dynamic coupling between the stress sensitivity and plastic strain rate. Specifically, the stress-driven term varies as a function of the logarithm of the strain rate, reflecting the progressive softening of the shear resistance with increasing deformation rates. In addition, an exponential correction term inversely dependent on $\log(\dot{\varepsilon})$ captures the nonlinear acceleration of shear deformation at low strain rates, providing a physically consistent representation of microstructural evolution mechanisms. These structural features endow the discovered model with enhanced flexibility in capturing shear-dominated deformation, particularly under dynamic loading conditions. The fitting performances of traditional empirical models and the data-driven discovered model are shown in Fig. 5D and Fig. 5E. It is evident that conventional models, which neglect strain rate effects, fail to accurately represent the rate-dependent behavior observed in lithium. In contrast, the discovered model achieves excellent predictive accuracy, yielding a relative error of only 0.533%. Furthermore, Fig. 5F illustrates the extrapolation capability of the model, confirming its robustness and reliability beyond the experimental range.

These findings suggest that data-driven modelling can also advance scientific understanding. By producing concise and accurate constitutive expressions, the framework enables a renewed examination of the underlying physical mechanisms. Unlike traditional empirical formulas, which typically reflect prior assumptions, AI-discovered models can reveal novel patterns beyond existing knowledge, offering new insights into material behavior.

**Discussion**

This work presents a graph-based equation discovery framework for the data-driven discovery of constitutive laws for solids, in which a directed graph is introduced to represent symbolic equations. In the graph, nodes encode operators and variables, edges represent the equation structure, and edge features describe the parametric dependencies. This framework enables a compact representation of mathematical expressions that can accommodate undetermined material-specific parameters and multisource datasets, making it particularly well-suited for the discovery of constitutive laws in scientific applications. Compared with phenomenological approaches, the proposed framework enables the autonomous discovery of compact, closed-form constitutive relationships directly from experimental data. It addresses the limitations of human intuition, which often constrain constitutive modelling to simple linear or low-order nonlinear forms, and allows for the identification of more expressive equations capable of capturing complex nonlinear behaviors with improved predictive accuracy. Unlike black-box machine learning models, the discovered expressions are fully interpretable while maintaining a high level of predictive performance. These explicit formulations can provide insights into underlying physical mechanisms and contribute to the advancement of scientific understanding. In addition, compared with conventional tree-based symbolic regression methods, graph-based representation yields more concise representations by utilizing node sharing, flexible edge connectivity, and edge features that encode parametric dependencies. This design enhances both representational capacity and optimization efficiency.

In the case studies with real-world experimental data, the discovered constitutive models achieve high predictive accuracy, physical interpretability, mathematical compactness and broad generalizability. They outperform classical empirical models both in fitting experimental data and in extrapolating across regimes of strain, strain rate, and temperature, demonstrating their robustness in capturing complex, coupled material responses. Notably, traditional equation discovery methods based on single-source data are prone to overfitting and often lack generalizability, particularly in the presence of sparse or noisy data. In practical settings, mechanical experimental data are typically scarce, and uncertainties may arise from variations in experimental procedures or environmental conditions. To address these challenges, the proposed framework incorporates material-specific parameters into edge features, enabling it to integrate data from multiple sources. This ensures that the discovered constitutive laws are generalizable and not overfitted to a specific material or condition. Consequently, the framework not only mitigates the limitations associated with insufficient single-source data but also improves the applicability of the discovered constitutive laws across a wide range of material systems and loading conditions.

Nevertheless, several limitations remain. The specification of graph templates introduces a form of prior structural bias, which may constrain the discovery process if not appropriately generalized. Moreover, the current implementation is limited to scalar constitutive responses. Extending the framework to accommodate tensorial quantities and to model anisotropic material behavior remains an open direction for future investigations. In addition, for complex problems involving multiple variables or coupled physical fields, further research is needed to balance representational expressiveness with computational efficiency. Nevertheless, by combining graph-based representation with symbolic discovery, this work contributes to a new paradigm of scientifically grounded machine learning for physical law discovery.

**Materials and methods**
**Graph representation of the equation.** In the proposed graph-based equation discovery framework, each candidate equation is encoded as a directed acyclic graph (DAG), denoted by $\mathcal{G} = (\mathcal{V}, \mathcal{E})$, where $\mathcal{V}$ is the set of nodes representing operators or variables, and $\mathcal{E}$ is the set of directed edges representing the computational dependencies between them. This graph-based representation enables flexible and compact expression of mathematical structures, surpassing traditional tree-based symbolic regression in terms of expressiveness and search efficiency. For node types and structure, $v \in \mathcal{V}$ belongs to one of the following categories: (1) Variable nodes, which represent input variables (e.g., strain rate $\dot{\varepsilon}$, temperature $T$) or the constant term "1"; and (2) operator nodes, which represent mathematical operations drawn from a predefined operator set which can be adjusted according to the physical scenarios. In this work, the operator set is defined as $\mathcal{O} = \{+, \times, \exp, \log, \text{pow}\}$, since the constitutive model typically does not involve trigonometric functions. Notably, our method is not restricted to these operators. It can be extended to incorporate additional operators, including trigonometric and inverse trigonometric functions, as well as hyperbolic functions. Furthermore, the framework supports the integration of user-defined operators, enabling extensibility for domain-specific applications. Each non-terminal node has an in-degree of 1, indicating a unique computational parent, whereas the out-degree is determined by the property of the operator. Specifically, *add* and *mul* have multiple out-degrees (>1), while other operators' in-degree is 1. Notably, although *log* and *exp* are typically treated as unary operators, their logarithmic base or exponent can be encoded into the corresponding edge features. As a result, these operators are represented with an out-degree of one in the graph structure.

In the graph, edges $s(v_i \to v_j) \in \mathcal{E}$ indicate that node $v_j$ uses the output of node $v_i$ as one of its inputs. Each edge is associated with an edge feature $e_{ij} \in \mathbb{R}$, which represents either a fixed scalar constant (e.g., 1, 0.5), or a learnable undetermined coefficient $C_i$ to be optimized during model evaluation. To avoid exhaustive search over the entire real number space, we define a discrete set of candidate constants for edge features: {-2, -1, -0.5, 0, 0.5, 1, 2, C}, which includes commonly occurring constant values. For other values not included in this set, the learnable undetermined coefficient that is later optimized according to the observation data is employed. Notably, the set of admissible edge features varies depending on the associated operator. For example, for the logarithmic operator, the edge feature set is restricted to {$e$, 10}, while for trigonometric functions, the corresponding set is {$\pi$, $2\pi$, 1, 2, -1, -2}. In this way, functions with undetermined coefficients, such as $x^{C_1}$ and $\log(C_1 x)$, can be represented by the graphs, which enriches the expressive ability of the proposed method. To parse a graph into an analytical expression, a post-order traversal is performed. Each node recursively computes its output based on the outputs of its children and the associated edge features.

**Graph templates.** To constrain the symbolic search space and ensure the mathematical validity of the generated equations, we introduced a structured mechanism, the graph template. A graph template defines a set of syntactic and topological constraints that govern how equations can be assembled from operators and variables within the graph-based representation. By restricting graph construction to a template-conforming space, we significantly improve the efficiency, robustness, and interpretability of the equation discovery process. Formally, a basic graph template is defined as $\mathcal{P} = (\mathcal{O}_{\text{valid}}, d_{max})$, where $\mathcal{O}_{\text{valid}} = \{add, mul, pow\}$ is the allowed set of operators, $d_{max}$=2 represents the maximum allowable graph depth. For the graph template of operators other than *add* and *mul*, their corresponding subgraphs are constrained to compositions of elements from the set *P*, which includes addition, multiplication, and division. This restriction effectively prevents the generation of overly complex terms and thereby improves the efficiency of the optimization process. During the graph generation process, candidate equations are constructed by sampling and assembling subgraphs that conform to the graph template. The process begins with the random selection of operators and corresponding graph templates, each of which is instantiated into subgraphs by combining them with randomly generated base structures from the basic graph template $\mathcal{P}$. The connections between these templates are subsequently randomly selected from *add* or *mul*, assembling them into a complete graph of equations.

**Graph evaluation.** Each generated graph in our framework represents a candidate symbolic equation that may include undetermined coefficients embedded as edge features. To quantitatively assess the quality of a candidate equation, we perform parameter identification followed by graph evaluation, which reflects the model's predictive accuracy with respect to observed data. Given a candidate graph $\mathcal{G}$ that corresponds to a symbolic expression $\hat{y} = f(x; \theta)$, where $\theta = \{C_1, C_2, ...C_n\}$ denotes the set of undetermined coefficients encoded in the graph's edge features, we optimize $\theta$ by minimizing the squared error loss between the model prediction and ground-truth data:

$$\mathcal{L}(\theta) = \frac{1}{N}\sum_{i=1}^{N}(f(x_i;\theta) - y_i)^2 \tag{8}$$

The optimization is performed using the L-BFGS algorithm(*51*), a quasi-Newton method suitable for smooth objective landscapes and small-to-moderate numbers of parameters. This optimization yields the best-fit parameters $\theta^*$ for the given graph. Notably, the initial values of unknown parameters may influence the final optimization outcome. To mitigate this sensitivity, we adopted a multi-start optimization strategy, which randomly generates 20 different initializations for the parameters of each candidate equation and retains the solution corresponding to the lowest post-optimization error as the final solution. For tasks involving a single experimental dataset, the graph's loss is defined as the loss obtained after parameter optimization: $\text{Loss}(\mathcal{G}) = \mathcal{L}(\theta^*)$. In cases where data from multiple sources (e.g., different materials) are used, we fit the parameters $\theta^{(j)*}$ independently for each dataset $D^{(j)} = \{(x_i^{(j)}, y_i^{(j)})\}_{i=1}^{N_j}$, and compute the average loss across all $M$ datasets: $\text{Loss}(\mathcal{G}) = \frac{1}{M}\sum_{j=1}^{M}\mathcal{L}^{(j)}(\theta^{(j)*})$. This multi-source evaluation strategy ensures that the discovered equations are generalizable and not overfit to a specific material or condition. Notably, unlike many equation discovery frameworks, we do not explicitly penalize model complexity in the loss function. Instead, model parsimony is implicitly enforced through structural constraints, such as the use of graph templates, subgraph deletion in the cross-over and bounded numbers of edges and unknown parameters. These designs prevent the generation of overly complex expressions, ensuring that the final equations remain both interpretable and computationally efficient, without sacrificing predictive performance.

**Graph cross-over and mutation.** To facilitate genetic programming for the discovery of equations by the proposed graph-based equation discovery framework, we designed the specific crossover and mutation processes for graphs. Graph crossover is implemented as a subgraph exchange mechanism between two parent graphs.

Specifically, a random node is selected in each parent graph, and the corresponding subgraphs rooted at these nodes are swapped. The resulting offspring graphs inherit structural components from both parents, enabling the recombination of functional motifs and promoting the emergence of new equation forms. Graph mutation operations introduce local stochastic perturbations to increase diversity and prevent premature convergence. Three distinct mutation strategies are employed. (1) Edge feature perturbation: A randomly selected edge is modified by replacing its associated feature with a new randomly sampled value. This allows the model to explore alternative parameterizations without altering the graph topology. (2) Subgraph replacement: A randomly chosen non-initial node is selected, and the subgraph rooted at this node is replaced with a newly generated subgraph sampled from the graph template. This operation enables the injection of structurally diverse functional forms while maintaining semantic consistency. (3) Subgraph deletion: A subgraph is randomly chosen and removed from the graph. The parent node is replaced by a constant term. This promotes structural simplification and serves as a form of regularization to prevent model overfitting. These operations collectively ensure a balance between exploration and exploitation in the graph-based search space, supporting the efficient discovery of compact and interpretable constitutive equations.

**The evolution of best equation.** The optimization process is driven by a graph-based genetic programming algorithm, designed to iteratively evolve a population of candidate graphs (equations) toward higher predictive accuracy. While the core structure of our evolutionary strategy follows the standard generate–evaluate–select–mutate paradigm of classical genetic algorithms, we incorporate a specialized extinction mechanism to avoid premature convergence and promote broader exploration of the solution space. The evolutionary process begins with the random generation of an initial population $S = \{\mathcal{G}_1, \mathcal{G}_2, \ldots, \mathcal{G}_N\}$, where each $\mathcal{G}$ is a graph representing a symbolic equation. Graphs are generated by sampling subgraphs from predefined graph templates, ensuring structural validity and diversity from the outset. At each generation, all the candidate graphs in the population are evaluated using the procedure described in the *Graph Evaluation* section. Each graph is assigned a loss value based on its prediction error after parameter optimization. Following evaluation, the population is ranked by the calculated loss (lower is better), and the top-performing individuals (typically the top 50%) are retained. These graphs are directly passed to the next generation, while the remaining individuals in the population are replaced by randomly generated graphs. To address the risk of stagnation in local optima, we incorporated a periodic extinction strategy. At predefined intervals (e.g., every five generations without improvement in population-best loss), extinction is triggered. Only the top *k* lowest-loss individuals are retained (*k*=5 in this work) and the remaining population is entirely replaced with new randomly generated graphs drawn from the template space. This mechanism reintroduces high diversity into the population and facilitates escape from local minima. In addition, the top five graphs from each generation are retained and carried over to the next generation until they are replaced by better-performing candidates. This strategy ensures that the best-performing equations are preserved throughout the evolutionary process and are not lost because of crossover and mutation. The evolutionary loop continues for a fixed number of generations (150 epochs in this work). The best-performing graphs in the final generation are returned as the finally discovered equations.

**Acknowledgements**

This work was supported and partially funded by the National Natural Science Foundation of China (Grant 52288101), the China Postdoctoral Science Foundation (Grant No. 2024M761535), the China National Postdoctoral Program for Innovative Talents (Grant No. BX20250063), and the Natural Science Foundation of Ningbo, China (No. 2023J027). This work is supported by the High Performance Computing Centers at Eastern Institute of Technology, Ningbo, and Ningbo Institute of Digital Twin.

**Author contributions**

H. X., and D.Z. conceived the idea, designed the study, and analyzed the results. H.X. developed the algorithm, performed the computations, and generated the results and figures. H.X., Y.C. and D.Z. wrote and edited the manuscript. D.Z. supervised the entire project.

**Declaration of interests**

The authors declare that they have no competing interests.